\documentclass[letterpaper,twocolumn,10pt]{article}
\usepackage{usenix,epsfig,endnotes}

\usepackage{xurl}
\usepackage{amsmath}

\usepackage{subfig}
\usepackage{graphicx}

\usepackage{bbding}

\usepackage{soul}
\usepackage{xcolor}
\usepackage{hyperref}
\usepackage{todonotes}

\begin{document}

%don't want date printed
\date{}

%make title bold and 14 pt font (Latex default is non-bold, 16 pt)
\title{\Large \bf CTI-HAL: A Human-Annotated Dataset for\\ Cyber Threat Intelligence Analysis}

%for single author (just remove % characters)
%\author{
%{\rm Your N.\ Here}\\
%Your Institution
%\and
%{\rm Second Name}\\
%Second Institution
% copy the following lines to add more authors
% \and
% {\rm Name}\\
%Name Institution
%} % end author

\author{
{\rm Sofia Della Penna, Roberto Natella, Vittorio Orbinato, Lorenzo Parracino, Luciano Pianese}\\
DIETI, Università degli Studi di Napoli Federico II, Naples, Italy\\
\{sofia.dellapenna, roberto.natella, vittorio.orbinato, lorenzo.parracino, luciano.pianese\}@unina.it\\
} 

\maketitle

% Use the following at camera-ready time to suppress page numbers.
% Comment it out when you first submit the paper for review.
\thispagestyle{empty}

\begin{abstract}
Organizations are increasingly targeted by Advanced Persistent Threats (APTs), which involve complex, multi-stage tactics and diverse techniques. Cyber Threat Intelligence (CTI) sources, such as incident reports and security blogs, provide valuable insights, but are often unstructured and in natural language, making it difficult to automatically extract information. 
Recent studies have explored the use of AI to perform automatic extraction from CTI data, leveraging existing CTI datasets for performance evaluation and fine-tuning. However, they present challenges and limitations that impact their effectiveness. To overcome these issues, we introduce a novel dataset manually constructed from CTI reports and structured according to the MITRE ATT\&CK framework.
To assess its quality, we conducted an inter-annotator agreement study using Krippendorff’s alpha, confirming its reliability. 
Furthermore, the dataset was used to evaluate a Large Language Model (LLM) in a real-world business context, showing promising generalizability.
\end{abstract}

% tabella related_work
\begin{table*}[ht]
\resizebox{\textwidth}{!}{%
\begin{tabular}{lcclcclc}
\hline
\textbf{Dataset}                                                            & \multicolumn{1}{l}{\textbf{Availability}} & \multicolumn{1}{l}{\textbf{Human-Labeled}} & \textbf{Granularity}                & \multicolumn{1}{l}{\textbf{Format}} & \multicolumn{1}{l}{\textbf{\# Techniques}} & \textbf{Source}                        & \multicolumn{1}{l}{\textbf{MITRE ATT\&CK}} \\ \hline
D-IT (Chen et al., 2024) \cite{CHEN2025104213}                              & \Checkmark \cite{AECR2025}                & \XSolidBrush                            & statement level                     & csv                                 & 201                                        & MITRE ATT\&CK KB                       & v14.1                                      \\ \hline
D-PE (Chen et al., 2024) \cite{CHEN2025104213}                              & \Checkmark \cite{AECR2025}                & N/A                                     & statement level                     & csv                                 & 189                                        & real CTI reports                       & v14.1                                      \\ \hline
TTPHunter [Sentence-Based] (Rani et al., 2023) \cite{TTPHunter}             & \Checkmark \cite{TTPHunterGit}            & \XSolidBrush                            & statement level                     & csv                                 & 50                                         & MITRE ATT\&CK KB                       & v13.1                                      \\ \hline
TTPHunter [Document-Based] (Rani et al., 2023) \cite{TTPHunter}             & \XSolidBrush                              & \Checkmark                              & document level                      & N/A                                 & N/A                                        & $50$ real CTI reports                       & v13.1                                      \\ \hline
TTPXHunter [Sentence-TTP] (Rani et al., 2024) \cite{TTPXHunter}             & \XSolidBrush                              & \XSolidBrush                            & statement level                     & N/A                                 & 193                                        & MITRE ATT\&CK KB                       & v15.1                                      \\ \hline
TTPXHunter [Report-TTP] (Rani et al., 2024) \cite{TTPXHunter}               & \XSolidBrush                              & \Checkmark                              & document level                      & N/A                                 & N/A                                        & 149 real CTI reports                       & v15.1                                      \\ \hline
CTI-to-MITRE with NLP (Orbinato et al., 2022)  \cite{orbinato2022automatic} & \Checkmark \cite{dessertlab_cti_to_mitre} & \XSolidBrush                            & statement level                     & csv                                 & 188                                        & MITRE ATT\&CK KB                       & v11.3                                      \\ \hline
rcATT (Legoy et al.,  2020) \cite{legoy2020rcatt_th}                        & \Checkmark \cite{legoy2020rcatt}          & \XSolidBrush                            & document level                      & csv                                 & 215                                        & MITRE ATT\&CK KB                       & v7.0                                       \\ \hline
MITREtrieval (Huang et al., 2024) \cite{MITREtrieval}                       & \Checkmark \cite{wmlab_MITREtrieval}      & N/A                                     & document level                      & json                                & 165                                        & mixed                                  & v10.1                                      \\ \hline
TRAM \cite{mitre2023tram}                                                   & \Checkmark \cite{mitre_tram_github}       & \Checkmark                              & statement level                     & json                                & 50                                         & real CTI reports                       & v13.1                                      \\ \hline
IntelEX Ground Truth (Xu et al., 2023) \cite{xu2023intelEX}                 & \Checkmark \cite{intelex11dataset}        & \Checkmark                              & document level                      & docx                                & 171                                        & 16 real CTI reports                       & v15.1                                      \\ \hline
LADDER (Alam et al., 2023) \cite{alam2023looking}                & \Checkmark \cite{ladder2023}              & \Checkmark                              & \multicolumn{1}{c}{statement level} & csv                                 & 31                                         & real CTI reports (Android malware) & v13.1                                      \\ \hline
LLMCloudHunter (Schwartz et al., 2024) \cite{schwartz2024llmcloudhunter}    & \XSolidBrush                              & \Checkmark                              & \multicolumn{1}{c}{N/A}             & N/A                                 & N/A                                        & 12 real CTI reports (cloud)           & v15.1                                      \\ \hline
aCTIon (Siracusano et al., 2023) \cite{siracusano2023time}                  & \XSolidBrush                              & \Checkmark                              & \multicolumn{1}{c}{N/A}             & N/A                                 & N/A                                        & 204 real CTI reports                       & v13.1                                      \\ \hline
AttacKG+ (Zhang et al., 2024) \cite{zhang2024attackgk}                      & \XSolidBrush                              & \Checkmark                              & \multicolumn{1}{c}{N/A}             & N/A                                 & N/A                                        & real CTI reports                       & v14.1                                      \\ \hline
CTI-ATE (Alam et al., 2024) \cite{alam2024ctibench}                        & \Checkmark \cite{ctibench2024}            & \XSolidBrush                            & \multicolumn{1}{c}{statement level} & tsv                                 & 115                                        & MITRE ATT\&CK KB                       & v15.1                                      \\ \hline
\textbf{CTI-HAL}                                                        & \Checkmark                                & \Checkmark                              & \textbf{statement level}            & \textbf{json}                       & \textbf{116}                               & \textbf{81 real CTI reports}              & \textbf{v15.1}                             \\ \hline
\end{tabular}
}
\caption{CTI Dataset Overview}
\label{tab:dataset_overview}
\end{table*}

\section{Introduction}
The complexity of cyberattacks faced by organizations is constantly increasing. Today, companies must face increasingly sophisticated threats, commonly known as APTs. In this context, \textit{reactive} security strategies, which respond only after an attack has occurred, are no longer sufficient. As a result, there is a growing shift towards \textit{proactive} security strategies that aim to anticipate the moves of attackers and neutralize threats before they manifest.

To develop \textit{proactive} security strategies, it is essential to gather information on APTs through CTI sources. CTI provides detailed analyses that aid in preventing and responding to cyberattacks by identifying trends, patterns, and relationships across multiple data sources. This intelligence enables security teams to take targeted, data-driven actions to enhance their defense posture effectively \cite{ibm_threat_intelligence}.
CTI supports a variety of activities, including \textit{Adversary Emulation} and \textit{Red Teaming}, \textit{Threat Hunting}, \textit{Risk Assessment}, and \textit{Incident Response}, making it a cornerstone of modern cybersecurity strategies.

The information collected from CTI sources can be structured using frameworks like MITRE ATT\&CK \cite{mitre_attack}, a knowledge base designed to understand cyber threats by analyzing cybercriminal behaviors.
MITRE ATT\&CK provides a standardized language that facilitates information sharing among security teams. 
However, automatically transforming \textit{unstructured} CTI sources, written in natural language, into a \textit{structured} format for security processes remains a significant challenge.

Some studies have attempted to overcome this limitation, by extracting \textit{structured} CTI from \textit{unstructured} sources with NLP techniques, and more recently LLMs \cite{arazzi2023nlp} \cite{AECR2025} \cite{TTPXHunter}. 
Despite their potential, several challenges persist. A key issue is the tendency to misinterpret benign sentences as anomalous, incorrectly identifying TTPs (Tactics, Techniques, and Procedures) that are not actually present. Another significant challenge is the phenomenon of ``\textit{hallucinations}'', where models fabricate nonexistent facts or produce inappropriate information in an attempt to generate a response \cite{yao2024llmlieshallucinationsbugs}.

Most studies use CTI datasets to evaluate the overall performance of AI-based CTI analyzers and to fine-tune them. However, these datasets often have significant limitations, including limited availability, the use of \textit{document-level} rather than \textit{statement-level} granularity, which results in a loss of correspondence between individual sentences and the identified TTPs, and reliance on short and generic TTP descriptions from the MITRE ATT\&CK Knowledge Base \cite{mitre_attack} rather than detailed CTI reports written by cybersecurity analysts.

In this work, we address these challenges by conducting a detailed review of existing datasets and introducing \textbf{CTI-HAL} (\textbf{CTI Human-Annotated Labels}), a new CTI dataset\footnote{\url{https://github.com/dessertlab/CTI-HAL}}, well-suited for the evaluation of NLP techniques. Our dataset includes reports of various sizes, with annotations on specific statements, allowing the assessment of a model’s ability to identify only the most relevant parts. Furthermore, it ensures traceability, facilitates \textit{cross-referencing}, and maintains high-quality standards through multiple annotators and \textit{cross-validation}. As a result, this new CTI dataset represents a versatile tool for research and applications in cybersecurity.

% LESSONS LEARNED?
%We also applied ... (progetto aziendale)
%- rispetto a valori riportati in paper recenti, le prestazioni di LLM possono essere molto più basse \cite{X}
%- LLM struggle with long CTI documents...
%- se applichiamo LLM su documenti da commercial CTI feed, otteniamo risultati comparabili a LLM su nostro dataset (generalizzabilità promettente)

We applied our dataset in a real-world business context, using it to evaluate the performance of an automation flow, based on an LLM, to extract TTPs from unstructured commercial CTI feeds. The lessons learned from this application are as follows:
\begin{itemize}
    %\item Data availability and variability are crucial. Availability allows experiments to be reproduced and results to be properly evaluated. Variability, both in terms of the topics covered and the size of the reports, is essential to cover a wider range of real-world scenarios.

    %\item Variability in topics and report size is crucial for generalizing across a wider range of real-world scenarios.

    \item Datasets have a significant impact on the accuracy of LLM under study. Compared to previous studies, our analysis found large differences, depending on the size and topics of CTI reports.

    \item The evaluated LLM struggles more with analyzing large CTI reports, whereas when analyzing concise reports, it achieves significantly better performance.
    
    \item When applying the LLM to commercial CTI feeds, we obtain results comparable to those achieved with our dataset, demonstrating promising generalizability.
\end{itemize}

This paper is structured as follows. Section \ref{sec:Related_Works} reviews related work, providing an overview of existing approaches and their limitations. Section \ref{sec:Methodology} describes the approach used to create the dataset, including a classification summary of the collected data. Section \ref{sec:quality_assessment} focuses on the quality assessment of the dataset, while Section \ref{sec:Evaluation} presents the application of the dataset in a real-world business context along with the evaluation results. Finally, Section \ref{sec:Conclusion} concludes the paper with a discussion of findings and future research directions.

% immagine workflow
\begin{figure*}[ht]
    \centering
    \includegraphics[width=\linewidth]{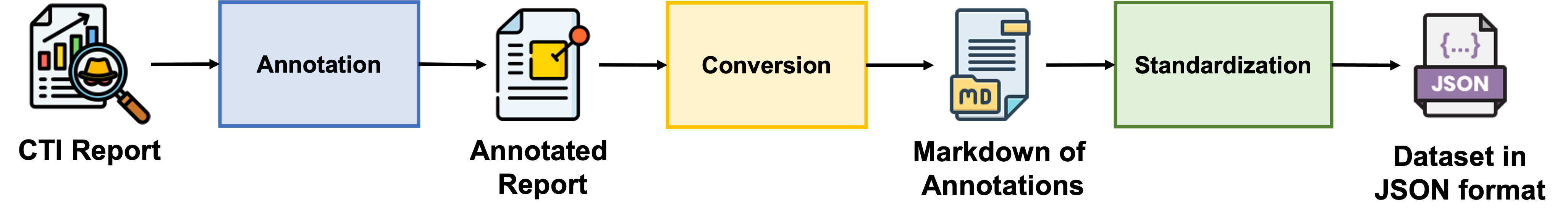}
    \caption{Workflow}
    \label{fig:flowchart}
\end{figure*}

\section{Related Work}\label{sec:Related_Works}

Several datasets have been proposed in the literature to assist in training and evaluating models for the extraction of TTPs from CTI reports. Table \ref{tab:dataset_overview} presents an overview of these datasets. 
%
%It is important to note that some of the considerations are valid at the time of writing this paper.

\textbf{Availability}. Publicly accessible datasets are essential to foster reproducibility, enable comparative evaluations, and support advancements in the field. 
However, some studies did not release their datasets at the time of writing this paper \cite{TTPHunter} \cite{TTPXHunter} \cite{schwartz2024llmcloudhunter} \cite{siracusano2023time} \cite{zhang2024attackgk}, limiting the ability to reproduce and validate their findings, and hindering further research.

\textbf{Human-Labeled}. The use of human-annotated datasets is crucial, particularly in the context of LLMs \cite{chatgpt_label} \cite{chatgpt_vs_human}. Human-labeled data ensures a higher degree of reliability compared to automated approaches. 
Some studies bypass this step by directly referencing the MITRE ATT\&CK Knowledge Base (KB) \cite{CHEN2025104213} \cite{TTPHunter} \cite{TTPXHunter} \cite{orbinato2022automatic} \cite{legoy2020rcatt_th} \cite{ctibench2024}. 
The KB provides short paragraphs to describe examples of attack techniques, tools, and campaigns \cite{mitre_attack}. Therefore, these studies only rely on brief technique descriptions, but do not consider the wider context. In practice, real CTI reports provide much larger descriptions that include details on the adversary’s motivations and operational context. As a result, relying solely on the KB’s predefined descriptions may lead to oversimplified threat representations that do not capture the full complexity of real-world cyber threats.

Others include human annotations on real CTI reports, often involving cybersecurity experts or multiple annotators to enhance validation \cite{TTPHunter} \cite{TTPXHunter} \cite{mitre2023tram} \cite{xu2023intelEX} \cite{alam2023looking} \cite{schwartz2024llmcloudhunter} \cite{siracusano2023time} \cite{zhang2024attackgk}. In these cases, real CTI reports are used as the primary source for annotation, sometimes focusing on specific cybersecurity topics, such as on Android and cloud attack techniques \cite{alam2023looking} \cite{schwartz2024llmcloudhunter}.
Nonetheless, none of these studies provide an \textit{inter-annotator agreement} evaluation to assess the quality of the annotations, leaving a gap in evaluating the consistency and reliability of their datasets.

\textbf{Granularity}. The level of granularity in dataset annotations plays a critical role. Using a \textit{statement-level} approach \cite{CHEN2025104213} \cite{TTPHunter} \cite{TTPXHunter} \cite{orbinato2022automatic} \cite{mitre2023tram} \cite{alam2023looking} \cite{alam2024ctibench}, rather than a \textit{document-level} one \cite{TTPHunter} \cite{TTPXHunter} \cite{legoy2020rcatt_th} \cite{MITREtrieval} \cite{xu2023intelEX}, significantly impacts the precision of CTI analysis.  
\textit{Document-level} granularity maps a list of TTPs to a group of CTI reports, without explicitly tracking which specific sentences contain TTP-relevant information. As a result, this approach lacks the necessary contextual linkage between the extracted techniques and their textual evidence, reducing the interpretability.  
In contrast, \textit{statement-level} granularity ensures that each identified technique is explicitly linked to its corresponding textual evidence. This \textit{bidirectional traceability} enhances interpretability and allows for more precise evaluations.  
For the remainder of this work, the terms \textit{statement} and \textit{sentence} will be used interchangeably.

Considering the limitations identified in previous studies, our work aims to address these gaps.
Our dataset is publicly available in JSON format and covers 116 techniques. The CTI reports used for its creation span various sectors, ensuring broader coverage of real-world scenarios. These reports also cover multiple APT groups, providing a diverse and comprehensive representation of adversary behaviors. 
Two annotators analyzed the reports to identify the TTPs within them, and the results were validated using inter-annotator agreement techniques. 
The annotations are performed at the sentence level, enabling a fine-grained mapping of techniques.

% immagine esempio metodologia
\begin{figure*}[ht]
     \centering
     \subfloat[{CTI Report}\label{fig:ex_annotation}]{\includegraphics[height=.2\textwidth]{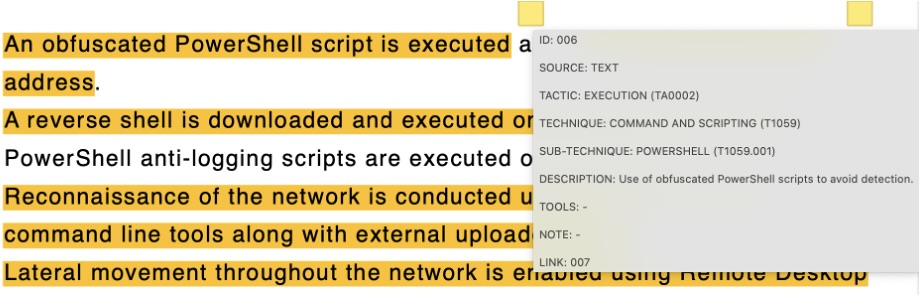}}
     \hfill
     \subfloat[{Markdown format}\label{fig:ex_annotation_md}]
     {\includegraphics[height=.25\textwidth]{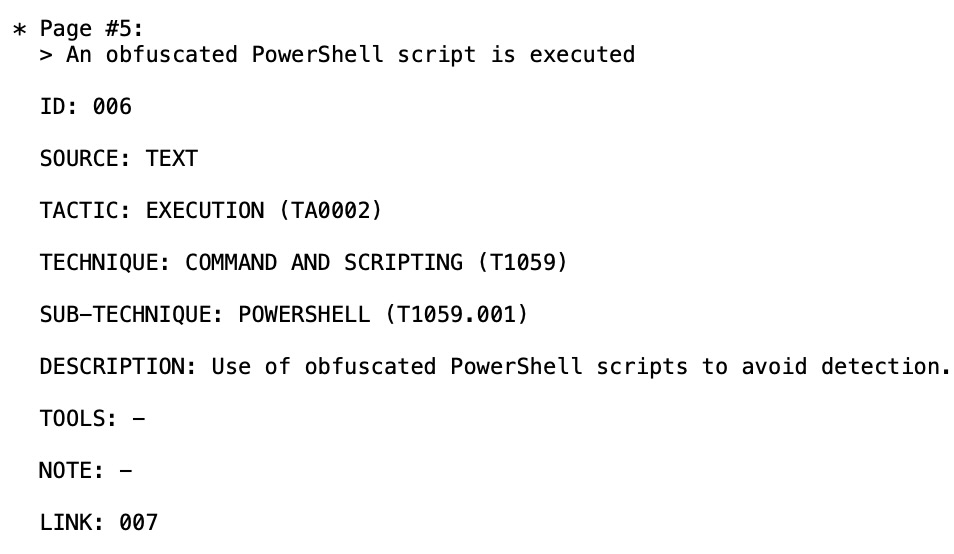}}
     \subfloat[{JSON format}\label{fig:ex_annotation_json}]{\includegraphics[height=.25\textwidth]{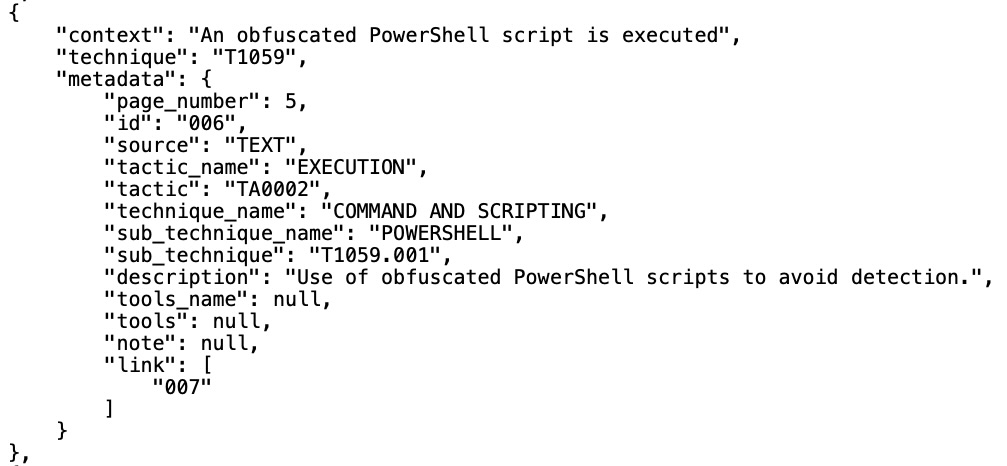}}
    \caption{Example of the application of the workflow}
    \label{fig:annotation}
\end{figure*}

% immagine classification_summary
\begin{figure*}[ht]
     \centering
     \subfloat[{Techniques}\label{fig:technique_occurrences}]{\includegraphics[height=.325\textwidth]{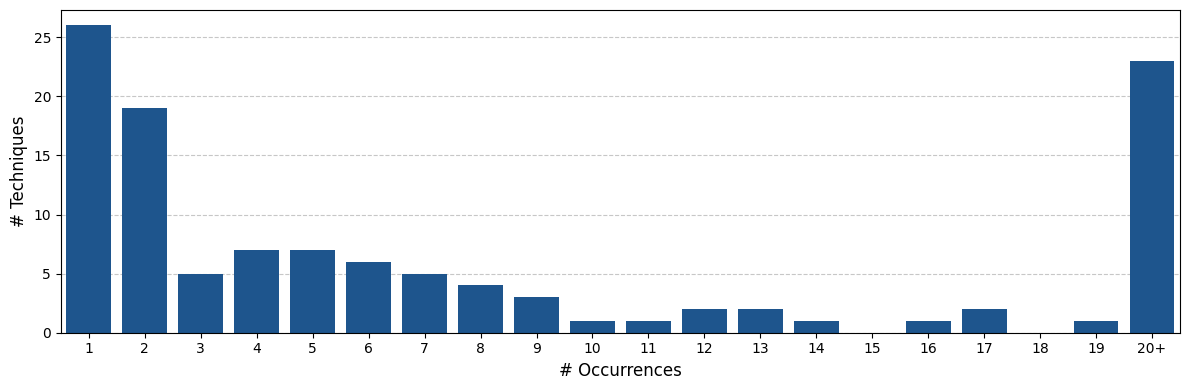}}
     \hfill
     \subfloat[{Tools}\label{fig:tools_occurrences}]
     {\includegraphics[height=.242\textwidth]{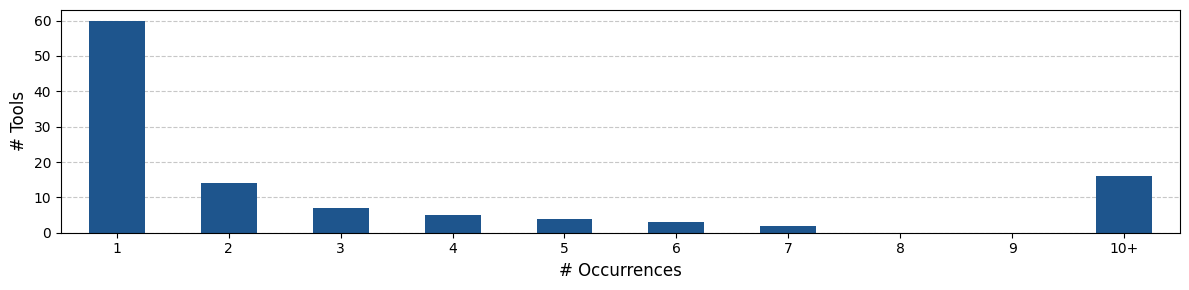}}
    \caption{Distribution of techniques and tools by number of occurrences.}
    \label{fig:occurrences}
\end{figure*}

\section{Methodology}\label{sec:Methodology}

We built CTI-HAL by carefully analyzing and manually annotating real CTI reports  written in natural language.
This approach is essential for creating an accurate and accessible dataset, providing a valuable resource to support threat intelligence activities.

The datasets discussed in Section \ref{sec:Related_Works} present some limitations. Some focus only on text fragments, failing to preserve the context of the attack described in the report, while others adopt a \textit{document-level} granularity, associating an entire document with a list of TTPs without specifying the text portions where they are mentioned.
These limitations hinder the application of AI techniques.

To overcome these limitations, we chose to directly annotate the documents using a \textit{statement-level} approach. Each sentence containing information related to a TTP was associated with the corresponding technique ID from the MITRE ATT\&CK framework, along with auxiliary information.
This approach also allowed us to maintain \textit{bidirectional traceability} between text fragments in the dataset and their specific locations in the original document.

The source of reports used to create the dataset is the Adversary Emulation Library \cite{ctid_adversary_emulation_library}, an open-source library that provides emulation plans for some attack campaigns, designed to allow organizations to test and evaluate their defensive capabilities.
Among the documents provided by the library, we analyzed those related to the following APTs: \textit{APT29} \cite{apt29_emulation_library} \cite{mitre_apt29}, \textit{Carbanak} \cite{carbanak_emulation_library} \cite{mitre_carbanak}, \textit{FIN6} \cite{fin6_emulation_library} \cite{mitre_fin6}, \textit{FIN7} \cite{fin7_emulation_library} \cite{mitre_fin7}, \textit{OilRig} \cite{oilrig_emulation_library} \cite{mitre_oilrig}, \textit{Sandworm} \cite{sandworm_emulation_library} \cite{mitre_sandworm}, and \textit{WizardSpider} \cite{wizardspider_emulation_library} \cite{mitre_wizardspider}.
We analyzed a total of 81 real CTI reports, excluding the ones that were unavailable.

The adopted approach involves analyzing CTI reports to identify sentences containing references to one or more of the MITRE ATT\&CK TTPs. 
To identify the portions to annotate, we focused on technical terms related to key elements such as known \textit{attack techniques}, specific types of \textit{malware}, names of \textit{tools} used by attackers, and the presence of particular \textit{file extensions}, which can indicate malicious files or links to malware. These aspects were identified based on the previous study of the MITRE ATT\&CK framework and through queries on the MITRE ATT\&CK KB. This approach enabled us to more accurately identify the relevant text portions for annotation, ensuring greater precision and consistency in the extracted data.

The workflow consists of three steps, shown in Figure \ref{fig:flowchart}.
These steps are described in detail below:
\begin{enumerate}
    \item \textbf{Annotation}: we manually analyzed the CTI report in PDF format identifying the sentences that reference TTPs. For each of these sentences, we associated the corresponding MITRE ATT\&CK technique ID. This information is reported as an annotation within the PDF.

    \item \textbf{Conversion}: the annotated PDF is automatically converted to a Markdown file. The file contains the text portions of the document along with their respective annotations. Each \textit{text-annotation pair} is assigned an identifying token to facilitate parsing in the subsequent step.
    
    \item \textbf{Standardization}: the data in the Markdown files is processed through a transformation using Python scripts, resulting in a dataset composed of JSON files, with each file representing a report.

\end{enumerate}

The described workflow was followed by two independent annotators, each working on separate documents. A subset of documents was used to evaluate the quality of the annotations, as detailed in Section \ref{sec:quality_assessment}.

The dataset creation process took eight weeks, with periodic meetings held every two weeks to monitor and discuss progress. 

\subsection{Example}

We present an example of the application of the described approach.
In particular, we consider the report: ``\textit{Big Game Hunting with Ryuk: Another Lucrative Targeted Ransomware}'' \cite{crowdstrike_ryuk}.

In the \textit{Annotation} phase, an annotation is added to each sentence in the document that contains TTP-related information.

An example is shown in Figure \ref{fig:ex_annotation}, describing the execution of a PowerShell script. Once executed, the script connects to a remote IP address, downloads a reverse shell, and executes it on the compromised host. For this reason, one of the identified techniques is \textit{T1059.001} (\textit{Command and Scripting Interpreter: PowerShell}) \cite{mitre_t1059_001}.

Each annotation consists of the following elements:
\begin{itemize}
    \item \textbf{ID}: distinguishes each annotation within the document.
    \item \textbf{SOURCE}: specifies whether the annotation refers to a text or an image.
    \item \textbf{TACTIC}: the tactic identified.
    \item \textbf{TECHNIQUE}: the technique identified.
    \item \textbf{SUB-TECHNIQUE}: the sub-technique identified.
    \item \textbf{DESCRIPTION}: description of the highlighted text and the identified technique.
    \item \textbf{TOOL}: any tools used, whether malicious software developed by APTs for specific purposes or red teaming toolkits employed in attacks.
    \item \textbf{NOTES}: any extra details to preserve the context of the document.
    \item \textbf{LINK}: link to other annotations enables tracking the sequence of techniques employed by the attacker.
\end{itemize}

In the \textit{Conversion} phase, the annotated document is transformed into a Markdown file using \textit{pdfannots} \cite{pdfannots_github}. Figure \ref{fig:ex_annotation_md} shows how an annotation appears in Markdown format.

In the \textit{Standardization} phase, we used Python scripts to organize the collected data, enabling the creation of the dataset of annotations in JSON format. Figure \ref{fig:ex_annotation_json} shows the structure of the JSON file for an annotation, which includes the following entries:
\begin{itemize}
    \item \textbf{CONTEXT}: The highlighted anomalous sentence.
    \item \textbf{TECHNIQUE}: The associated technique.
    \item \textbf{METADATA}: All additional information, including \textit{page\_number}, \textit{id}, \textit{source}, \textit{tactic}, \textit{sub-technique}, \textit{description}, \textit{tool}, \textit{notes}, and \textit{link}.
\end{itemize}

\subsection{Classification Summary}
After completing the dataset creation process, it is possible to summarize the classification results.
Table \ref{tab:avg_sentences_techniques} presents, for each APT, the number of documents analyzed, the average number of annotated sentences per document, and the average number of techniques identified per document. 
The dataset covers both nation-state APT groups and cybercriminal organizations, spanning various sectors such as energy, finance, telecommunications, intelligence, and government. This diversity ensures a comprehensive representation of threats across different domains, enhancing the dataset's applicability to real-world cybersecurity scenarios.

\begin{table}[ht]
\resizebox{\columnwidth}{!}{%
\begin{tabular}{|c|c|c|c|}
\hline
\textbf{APT}   & \textbf{\# Docs} & \textbf{\# Sentences (avg)} & \textbf{\# Techniques (avg)} \\ \hline
APT29 (L) & 12                           & 24                                   & 21                                     \\ \hline
APT29 (S) & 12                           & 24                                   & 24                                     \\ \hline
Carbanak            & 10                           & 29                                   & 27                                     \\ \hline
FIN6                & 11                           & 25                                   & 23                                     \\ \hline
FIN7                & 18                           & 15                                   & 14                                     \\ \hline
OilRig              & 8                            & 22                                   & 22                                     \\ \hline
Sandworm            & 7                            & 12                                   & 11                                     \\ \hline
WizardSpider        & 3                            & 21                                   & 20                                     \\ \hline
\end{tabular}%
}
\caption{Average sentences and techniques per APT}
\label{tab:avg_sentences_techniques}
\end{table}

The dataset includes a total of $116$ \textit{techniques}, $104$ \textit{sub-techniques}, and $111$ \textit{tools}.
To provide further insights into the dataset, we present a histogram where each bar represents the number of techniques that appear a specific number of times [Figure \ref{fig:technique_occurrences}]. The \textit{i}-th bar indicates how many techniques occur \textit{i} times in the dataset, while the “$20+$” column aggregates the count of techniques that appear more than $20$ times.
From the histogram, it is evident that many techniques appear only once, highlighting their more specialized and specific nature. However, there are also numerous techniques that appear more than $20$ times, indicating that certain techniques are more common and frequently used across multiple attacks.
Among those that appear more than 20 times, some of the most common techniques include \textit{T1059} (\textit{Command and Scripting Interpreter}), where adversaries may abuse interpreters to execute commands or scripts \cite{mitre_t1059}; \textit{T1566} (\textit{Phishing}), where adversaries may send phishing messages to gain access to victim systems \cite{mitre_t1566}; \textit{T1027} (\textit{Obfuscate Files or Information}), where adversaries may attempt to make a file difficult to discover or analyze by obfuscating its contents on the system \cite{mitre_t1027}; \textit{T1105} (\textit{Ingress Tool Transfer}), where adversaries may transfer tools or other files from an external system into a compromised environment \cite{mitre_t1105}; and \textit{T1071} (\textit{Application Layer Protocol}), where adversaries may communicate using application layer protocols to avoid detection or network filtering by blending in with existing traffic \cite{mitre_t1071}.

We also present the same type of histogram to represent the distribution of identified tools based on the number of occurrences [Figure \ref{fig:tools_occurrences}]. In this case, the “$10+$” column aggregates the count of tools that appear more than 10 times.
In this case, many tools appear only once, highlighting the tendency of APTs to use specific tools for targeted actions.
There are also tools that are used more extensively, as well as tools that are specific to certain APTs, which appear multiple times within the same document.
Among those that appear more than 10 times, some of the most common tools include \textit{S0030} (\textit{Carbanak}), a full-featured remote backdoor intended for espionage, data exfiltration, and providing remote access to infected machines \cite{mitre_s0030}; \textit{S0046} (\textit{CozyDuke}), a modular malware platform whose backdoor component can be instructed to download and execute a variety of modules with different functionality \cite{mitre_s0046}; \textit{S0050} (\textit{CosmicDuke}), a malware \cite{mitre_s0050} that collects information from the infected host and exfiltrates it to a C2 server; \textit{S0154} (\textit{Cobalt Strike}), a commercial, full-featured remote access tool \cite{mitre_s0154}; and \textit{S0053} (\textit{SeaDuke}), a malware \cite{mitre_s0053} used as a secondary backdoor for victims already compromised by \textit{CozyDuke}.

In conclusion, our dataset encompasses a wide range of \textit{techniques} and \textit{tools}. Analyzing the most common \textit{techniques} and \textit{tools} reveals distinct patterns in how attacks are conducted.
For example, attackers often use phishing to gain access to victim hosts, after which they transfer tools or scripts, collect sensitive information, and exfiltrate it.
These operations can be performed either with custom tools developed by the attackers or by using red teaming toolkits like \textit{Cobalt Strike} \cite{mitre_s0154}.

% tabella valori di alpha
\begin{table*}[ht]
\centering
\begin{tabular}{|p{0.15\linewidth}|p{0.25\linewidth}|p{0.50\linewidth}|}
\hline
\textbf{Value of $\alpha$} & \textbf{Interpretation} & \textbf{Details} \\ \hline
$\alpha<0$                  & \textit{Systematic Disagreement}   & There is a systematic disagreement between annotators, greater than what would be expected by chance.                     \\ \hline
$0 \leq \alpha < 0.2$       & \textit{Poor Agreement}            & The agreement is minimal and not better than what could be expected by chance.               \\ \hline
$0.2 \leq \alpha < 0.4$     & \textit{Fair Agreement}            & The agreement is slightly better than chance but still insufficient for many practical applications. \\ \hline
$0.4 \leq \alpha < 0.6$     & \textit{Moderate Agreement}        & The agreement is acceptable in some contexts but often requires improvement.  \\ \hline
$0.6 \leq \alpha < 0.8$     & \textit{Substantial Agreement}     & The agreement is considered good for most applications.                             \\ \hline
$0.8 \leq \alpha \leq 1$    & \textit{Perfect Agreement}         & There is a high level of agreement between annotators.        \\ \hline
\end{tabular}
\caption{Classification of Krippendorff's $\alpha$ values}
\label{tab:alpha_value}
\end{table*}

\section{Quality Assessment}\label{sec:quality_assessment}

Most of the datasets discussed in Section \ref{sec:Related_Works} lack validation of the quality of their annotations, which can lead to unreliable results.

To ensure the \textit{reliability} of the dataset, we emphasized the importance of its quality. We employed the \textit{inter-annotator agreement} technique, where two independent annotators followed the same workflow described in Section \ref{sec:Methodology} on the same CTI reports. Since the annotations were performed manually, they are prone to human errors.

We conducted \textit{quality assessment} using reports related to APT29 \cite{mitre_apt29} \cite{apt29_emulation_library}. To assess the degree of agreement between the two annotators, referred to here as \textit{L} and \textit{S}, we used \textit{Krippendorff’s Alpha coefficient}, a measure of inter-annotator reliability that tests the agreement between annotators on categorical, ordinal, or nominal data.

\subsection{Krippendorff's Alpha coefficient}
Krippendorff's alpha is a measure of inter-annotator reliability used to determine the \textit{level of agreement} between two or more annotators \cite{krippendorff2004content}.

The formula to calculate it is shown in Equation \ref{eq:alpha}.
\begin{equation}\
    \alpha=1-\frac{D_o}{D_e}
    \label{eq:alpha}
\end{equation}
Where:
\begin{itemize}
    \item \(D_o\) (\textit{Observed Discordance}): the proportion of discordant annotations among the annotators.
    \item \(D_e\) (\textit{Expected Discordance}): the discordance that would be expected if the annotations were independent.
\end{itemize}

The \textit{Observed Discordance} (\(D_o\)) \cite{krippendorff2013reliability} is the proportion of disagreements between annotators, counting the number of times annotators choose different classifications out of the total number of annotations (Equation \ref{eq:do}).
\begin{equation}
    D_o=\frac{\text{No. of discordances}}{\text{No. of annotations}}
    \label{eq:do}
\end{equation}

The \textit{Expected Discordance} (\(D_e\)) \cite{krippendorff2013reliability} is based on the probability that two annotators choose the same category randomly. Its calculation relies on the relative frequency of the annotated categories, estimating how often they would coincide by chance (Equation \ref{eq:de}).
\begin{equation}
    D_e=1-P_e
    \label{eq:de}
\end{equation}
where \(P_e\) represents the \textit{Expected Likelihood of Agreement}, calculated as the sum of the products of the relative likelihood of each category assigned by the two annotators. The relative probabilities reflect how often annotators attribute a specific category during the annotation process.

The values of Krippendorff's alpha vary between $-1$ and $1$: a value of $1$ indicates unanimous agreement among the annotators, $0$ suggests that classifications occur by chance and negative values indicate that the annotators disagree.
Table \ref{tab:alpha_value} presents the classification of alpha values and their interpretation.

\subsection{Similarity Metrics}
When identifying a sentence with relevant informational content, annotators may not highlight exactly the same portion of text. This can lead to one annotator selecting only a subsection of what the other has highlighted or to variations in the number of words chosen. 
To analyze these overlaps in the highlighted text more accurately, we adopted the following textual \textit{Similarity Metrics}:
\begin{itemize}
    \item \textit{Sequence Matcher} \cite{python_difflib_sequenceMatcher}: measures the syntactic similarity between two strings by comparing their literal content.
    \item \textit{BLEU} \cite{acl_p02_1040}: assesses similarity based on \textit{n-gram} precision, that is how closely a sequence of words in a target sentence matches a reference sentence.
    \item \textit{Embedding Distance}: measures the similarity between two texts using word vector representations called word embeddings, through SpaCy \cite{spacy}. SpaCy's \textit{en\_core\_web\_sm} model \cite{spacy_model_en} was chosen because it represents an ideal compromise between efficiency and accuracy, especially for short texts. 
\end{itemize}

Each pair of annotations evaluated using similarity metrics is assigned a score that reflects the degree of correspondence between the highlighted texts. A low score indicates poor similarity between the two text portions, while a high score indicates strong similarity. 
To ensure that annotation pairs refer to the same portion of text, we applied an \textit{acceptance threshold}: annotations that do not meet this value are discarded, as they are considered irrelevant.
In some cases, a sentence identified by one annotator could be associated with multiple sentences identified by the other, each with different scores.
To ensure that only the most relevant matches were considered, we implemented a filter to select annotation pairs with the highest similarity score, ensuring that only the most consistent ones were included.

\subsection{Results}
The analysis described in the previous section was applied to all CTI reports from the Adversary Emulation Library related to APT29.

\begin{figure}[ht]
    \centering
    \includegraphics[width=\linewidth]{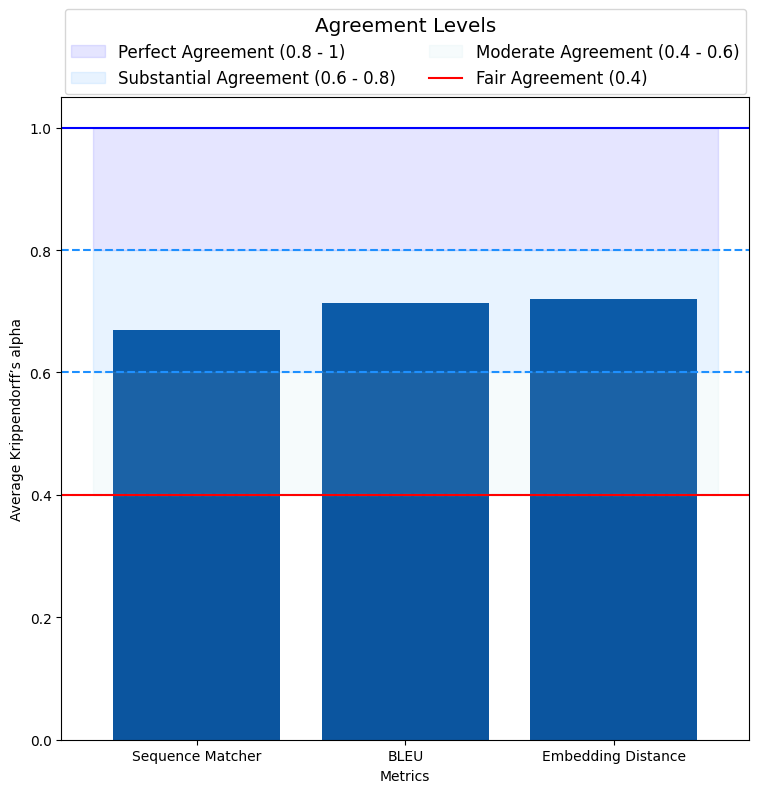}
    \caption{APT29 - Krippendorff's $\alpha$}
    \label{fig:average_alpha}
\end{figure}

The average Krippendorff’s alpha values for each similarity metric are presented in Figure \ref{fig:average_alpha}.
The overall average, calculated as the mean of the average values from the three metrics, is $0.70$. This indicates a \textit{Substantial Agreement} between the annotators.
This result suggests that, despite some variations in individual annotations, there is significant consistency in the interpretations of the documents.
These findings confirm the reliability of the collected information and, consequently, the quality of the dataset, providing a solid foundation for further studies and analyses.

For the sake of clarity, we provide an example of a comparison between annotations for the document ``\textit{Not So Cozy: An Uncomfortable Examination of a Suspected APT29 Phishing Campaign}'' \cite{google2025_apt29_phishing}, which was independently analyzed by both annotators, as previously detailed.
For each metric, the number of annotations that share the same highlighted text, referred to as '\textit{Common annotations}', and those that match both the text and the identified technique, referred to as '\textit{Concordant annotations}' are reported.
In this case, annotator L produced $22$ annotations, while annotator S produced $19$. 
Table \ref{tab:alpha_document_seaduke} presents the number of common and concordant annotations identified using each similarity metric, along with the corresponding Krippendorff’s alpha value.

\begin{table}[ht]
\centering
\resizebox{\columnwidth}{!}{%
\begin{tabular}{|l|c|c|c|c|}
\hline
{\textbf{Metric}} & \textbf{Common} & \textbf{Concordant} & $\alpha$ & \textbf{Agreement Level} \\ \hline
Sequence Matcher                           & $11$              & $8$                & $0.67$              & \textit{Substantial}   \\ \hline
BLEU                                        & $12$              & $9$                & $0.69$              & \textit{Substantial}   \\ \hline
Embedding Distance                                      & $9$              & $6$                 & $0.60$             & \textit{Substantial}   \\ \hline
\end{tabular}%
}
\caption{Krippendorff’s $\alpha$ for the Example Document}
\label{tab:alpha_document_seaduke}
\end{table}

The average Krippendorff’s alpha value is $0.65$, indicating a \textit{Substantial Agreement} between the annotators.
\section{Evaluation}\label{sec:Evaluation}
Companies are often targeted by malicious actors with different objectives and characteristics (e.g., cybercriminals, hacktivists, and industrial espionage). Since these actors have distinct purposes, it is challenging for companies to defend themselves effectively.
One of the main goals of companies is to recognize TTPs from targeted commercial feeds to develop appropriate defenses.

We collaborated with a large enterprise in the logistics domain\footnote{Anonymous for confidentiality reasons.} to analyze APTs specific to that sector. The company implemented an automation flow, based on an LLM, to structure the information from commercial CTI feeds. The model used was \textit{Claude 3 Haiku} (\texttt{anthropic.claude-3-haiku-20240307-v1:0}) \cite{anthropic2024claude3}, which was instructed to detect TTPs in CTI reports. 
Our dataset was used as ground truth to evaluate the model.

In the case study, CTI reports are written in natural language, are transmitted by email in PDF format, and include IoCs. 
These data are first normalized into a standard format and then analyzed through the following stages:
\begin{enumerate}
    \item \textbf{Email Parsing}: CTI data are extracted from the email.
    \item \textbf{IoC Analysis}: IoCs are analyzed to obtain a list of hashes, IP addresses, and domains.
    \item \textbf{PDF Analysis}: using \textit{Claude 3 Haiku} \cite{anthropic2024claude3}, the automation flow analyzes the reports and generates a JSON file for each report, with information on the identified attack techniques and related vulnerabilities, campaign names, involved sectors and nations.
    \item \textbf{Filtering}: the extracted data are filtered to retain only those relevant to the company’s sectors and nations.
\end{enumerate}

This automation flow was integrated into the company’s workflow and monitored for three months.
At the end of the observation phase, the collected data were analyzed, revealing that the most frequent attack techniques in this corporate context are \textit{Phishing} (\textit{T1566}) \cite{mitre_t1566}, \textit{Command and Scripting Interpreter} (\textit{T1059}) \cite{mitre_t1059}, and \textit{Obfuscated Files or Information} (\textit{T1027}) \cite{mitre_t1027}.
These techniques are also among the most frequent in our dataset, highlighting its strong representativeness of real-world threats.
By analyzing the identified techniques, a report is generated that provides a set of mitigation strategies recommended by MITRE ATT\&CK to strengthen the company’s defenses against the detected attack techniques.

To evaluate the ability of \textit{Claude 3 Haiku} \cite{anthropic2024claude3} to detect techniques in CTI reports, we conducted three experiments in which the model was provided with different types of CTI reports.
For these experiments, we crafted a prompt. The structure of the prompt to be submitted to the LLM is crucial, as an inadequate structure could result in ambiguous, incorrect or imprecise responses.
The prompt used for the experiments combines several prompt engineering techniques \cite{deepset2024promptengineering}: \textit{role prompting} to assign the AI a role as a Threat Intelligence expert, \textit{zero-shot prompting} to respond without specific examples, and \textit{output formatting} to structure the responses in a JSON format.

To evaluate the ability of \textit{Claude 3 Haiku} \cite{anthropic2024claude3} to detect techniques in CTI reports, we conducted three experiments in which the model was provided with different types of CTI reports.  
For these experiments, we crafted a prompt. The structure of the prompt to be submitted to the LLM is crucial, as an inadequate structure could result in ambiguous, incorrect, or imprecise responses.  
The prompt used for the experiments combines several prompt engineering techniques \cite{deepset2024promptengineering}: \textit{role prompting} to assign the AI a role as a Threat Intelligence expert, \textit{zero-shot prompting} to respond without specific examples, and \textit{output formatting} to structure the responses in a JSON format.  
In particular, the prompt is designed to generate a structured output that includes the MITRE ATT\&CK technique code and its name, along with a motivation explaining why the technique was identified based on the provided text. This structured format ensures consistency and facilitates the analysis of the outputs.  

\subsection{Results}
\begin{figure}[ht]
    \centering
    \begin{minipage}{0.48\textwidth}
        \centering
        \includegraphics[width=\textwidth]{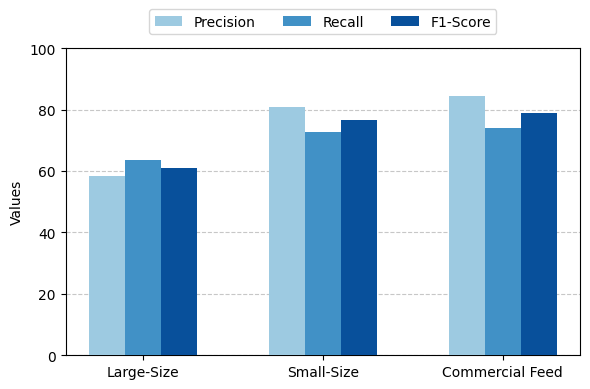}
    \end{minipage}
    \hfill
    \begin{minipage}{0.48\textwidth}
        \centering
        \resizebox{\textwidth}{!}{%
        \begin{tabular}{|l|c|c|c|}
        \hline
        \multicolumn{1}{|c|}{\textbf{Metrics}} & \multicolumn{1}{l|}{\textbf{Large-Size}} & \multicolumn{1}{l|}{\textbf{Small-Size}} & \textbf{Commercial Feed} \\ \hline
        \textit{Precision}                     & 58.55\%                                    & 80.95\%                                    & 84.37\%               \\ \hline
        \textit{Recall}                        & 63.75\%                                    & 72.64\%                                    & 73.97\%               \\ \hline
        \textit{F1-Score}                      & \textbf{61.04\%}                                    & \textbf{76.57\%}                                    & \textbf{78.83\%}               \\ \hline
        \end{tabular}
        }
    \end{minipage}
    \caption{Evaluation of CTI extraction}
    \label{fig:evaluation_results}
\end{figure}

We evaluated the performance of the model using several key metrics to assess its ability to generate accurate and relevant responses. These metrics include \textit{Precision}, \textit{Recall}, and \textit{F1-Score}. The evaluation was carried out by comparing the predictions of the model with the entries contained in our dataset.
The overall results for the three experiments are presented in Figure \ref{fig:evaluation_results}.

In \textit{Large-Size Report Evaluation}, we selected CTI reports from our dataset, specifically related to \textit{APT29} \cite{apt29_emulation_library}, \textit{CARBANAK} \cite{carbanak_emulation_library}, \textit{FIN6} \cite{fin6_emulation_library}, \textit{FIN7} \cite{fin7_emulation_library}, \textit{OILRIG} \cite{oilrig_emulation_library}, and \textit{WIZARDSPIDER} \cite{wizardspider_emulation_library}.
The selected reports are complex and detailed, with sizes ranging between $4$ KB ($\sim{400}$ words) and $20$ KB ($\sim{2000}$ words).
The goal of this experiment is to evaluate the performance of the model when analyzing large reports.

In the \textit{Small-Size Report Evaluation}, we selected a subset of reports from the initial experiment, ranging in size from $4$ KB ($\sim{400}$ words) to $8$ KB ($\sim{800}$ words). 
We made this choice because we hypothesize that report size may significantly influence the performance of the model, and also to align the report sizes with those typically found in commercial CTI feeds.

For \textit{Commercial Feed Evaluation}, we used commercial CTI feeds, with sizes typically ranging between $2$ KB ($\sim{200}$ words) and $8$ KB ($\sim{800}$ words).
This experiment evaluates the performance of the model on commercial CTI feeds.

The analysis of the results confirms the hypothesis that report size significantly affects the performance of the model. With \textit{Large-Size} reports, the model achieved an \textit{F1-score} of $61.04\%$, while with \textit{Small-Size} reports led to an improvement, reaching $76.57\%$, highlighting the effectiveness of LLMs in processing more concise information.
The model performed even better on \textit{commercial CTI feeds}, achieving an \textit{F1-score} of $78.83\%$.
This is because these sources present more compact information, fewer irrelevant sentences, and content that is more directly focused on TTPs compared to traditional CTI reports.
Moreover, the performance on commercial CTI feeds is comparable to that obtained with shorter CTI reports. This suggests that our dataset demonstrates \textit{promising generalizability}, as it effectively represents real-world business context.

To contextualize our results, we compare the performance metrics obtained in our study with those reported in previous work, which refer to their own datasets. Table \ref{tab:PM_comparison} presents a comparative analysis of results from other studies.

\begin{table}[ht]
\resizebox{\columnwidth}{!}{%
\centering
\begin{tabular}{|l|c|c|c|}
\hline
\textbf{Work}                                                            & \multicolumn{1}{l|}{\textbf{Precision}} & \multicolumn{1}{l|}{\textbf{Recall}} & \multicolumn{1}{l|}{\textbf{F1-Score}} \\ \hline
AECR (Chen et al., 2024) \cite{CHEN2025104213}                           & 64.1\%                                  & 68.3\%                               & 65.5\%                                 \\ \hline
TTPHunter (Rani et al., 2023) \cite{TTPHunter}                           & 74.0\%                                  & 77.0\%                               & 75.0\%                                 \\ \hline
TTPXHunter (Rani et al., 2024) \cite{TTPXHunter}                         & 97.4\%                                  & 96.2\%                               & 97.1\%                                 \\ \hline
rcATT (Legoy et al.,  2020) \cite{legoy2020rcatt_th}                     & 72.2\%                                  & 2.1\%                                & 4.0\%                                  \\ \hline
MITREtrieval (Huang et al., 2024) \cite{MITREtrieval}                    & 31.0\%                                  & 74.0\%                               & 43.7\%                                 \\ \hline
IntelEX (Xu et al., 2023) \cite{xu2023intelEX}                           & 69.6\%                                  & 75.6\%                               & 72.4\%                                 \\ \hline
LADDER (Alam et al., 2023) \cite{alam2023looking}                        & 65.0\%                                  & 63.0\%                               & 64.0\%                                 \\ \hline
LLMCloudHunter (Schwartz et al., 2024) \cite{schwartz2024llmcloudhunter} & 62.0\%                                  & 75.0\%                               & 68.0\%                                 \\ \hline
AttacKG+ (Zhang et al., 2024) \cite{zhang2024attackgk}                   & 54.5\%                                  & 58.8\%                               & 56.6\%                                 \\ \hline
CTI-Bench (Alam et al., 2024) \cite{alam2024ctibench}                    & N/A                                     & N/A                                  & 62.1\%                                 \\ \hline
\textit{\textbf{Large-Size Report Evaluation}}                           & \textbf{58.6\%}                         & \textbf{63.8\%}                      & \textbf{61.0\%}                        \\ \hline
\textit{\textbf{Small-Size Report Evaluation}}                           & \textbf{80.9\%}                         & \textbf{72.6\%}                      & \textbf{76.6\%}                        \\ \hline
\textit{\textbf{Commercial Feed Evaluation}}                             & \textbf{84.4\%}                         & \textbf{73.8\%}                      & \textbf{78.8\%}                        \\ \hline
\end{tabular}%
}
\caption{Comparison of CTI extraction across studies}
\label{tab:PM_comparison}
\end{table}

Some studies do not make their datasets available, making it impossible to reproduce the high performance reported in their results.

These results show a high variability of accuracy, depending on the dataset. Therefore, using a publicly-available dataset is important for reproducibility and comparability. Moreover, ensuring data variability is crucial in these evaluations, both in terms of the diversity of topics covered in the reports, and in terms of size. Some studies focus exclusively on reports from specific sectors, which can lead to models that are too specialized within a narrow domain, limiting their adaptability to a wider range of scenarios. The size of the reports, in particular, plays a significant role, as document length directly impacts the performance of LLMs.

Regarding our approach, the results demonstrate its ability to outperform several existing methods on small-sized reports. Additionally, when tested on commercial feeds, it shows strong effectiveness in handling structured data in operational contexts. Although performance on large reports is lower, it remains competitive with the best existing solutions, suggesting potential for further improvement.
\section{Conclusion}\label{sec:Conclusion}
This work introduces a novel CTI dataset that overcomes the limitations of existing datasets, providing a more comprehensive and structured resource for cybersecurity applications. 
We constructed the dataset from real-world CTI reports of varying sizes, ensuring its applicability to a wide range of attack scenarios, and based it on the MITRE ATT\&CK framework. The dataset maintains \textit{bidirectional traceability} between the original documents and the data, enhancing both transparency and accuracy. Additionally, we validated its quality through an \textit{inter-annotator agreement} study, confirming its \textit{reliability}.
Furthermore, the evaluation of an LLM in a real-world business context on this dataset highlights its \textit{promising generalizability}. 

CTI-HAL offers a valuable tool for advancing AI-driven cybersecurity solutions, enabling the development of more accurate and effective models.
\section*{Acknowledgments}
Sofia Della Penna and Lorenzo Parracino are both main authors of this work. 
We would like to thank Raffaele D’Ambrosio for the helpful discussions and support. 
This work has been partially supported by the \textit{IDA—Information Disorder Awareness} Project funded by the European Union-Next Generation EU within the SERICS Program through the MUR National Recovery and Resilience Plan under Grant PE00000014, and by project \emph{GENIO} (CUP B69J23005770005) funded by MIMIT.

{\footnotesize \bibliographystyle{IEEEtran}
\bibliography{references}}

\begin{thebibliography}{10}
\providecommand{\url}[1]{#1}
\csname url@samestyle\endcsname
\providecommand{\newblock}{\relax}
\providecommand{\bibinfo}[2]{#2}
\providecommand{\BIBentrySTDinterwordspacing}{\spaceskip=0pt\relax}
\providecommand{\BIBentryALTinterwordstretchfactor}{4}
\providecommand{\BIBentryALTinterwordspacing}{\spaceskip=\fontdimen2\font plus
\BIBentryALTinterwordstretchfactor\fontdimen3\font minus \fontdimen4\font\relax}
\providecommand{\BIBforeignlanguage}[2]{{%
\expandafter\ifx\csname l@#1\endcsname\relax
\typeout{** WARNING: IEEEtran.bst: No hyphenation pattern has been}%
\typeout{** loaded for the language `#1'. Using the pattern for}%
\typeout{** the default language instead.}%
\else
\language=\csname l@#1\endcsname
\fi
#2}}
\providecommand{\BIBdecl}{\relax}
\BIBdecl

\bibitem{CHEN2025104213}
M.~Chen, K.~Zhu, B.~Lu, D.~Li, Q.~Yuan, and Y.~Zhu, ``{AECR: Automatic attack technique intelligence extraction based on fine-tuned large language model},'' \emph{Computers \& Security}, 2024.

\bibitem{AECR2025}
M.~Chen, ``{AECR: Automatic attack technique intelligence extraction},'' \url {https://github.com/cmh14/AECR.git}, 2025, gitHub repository.

\bibitem{TTPHunter}
N.~Rani, B.~Saha, V.~Maurya, and S.~K. Shukla, ``Ttphunter: Automated extraction of actionable intelligence as ttps from narrative threat reports,'' in \emph{Proceedings of the 2023 Australasian Computer Science Week}, 2023, pp. 126--134.

\bibitem{TTPHunterGit}
{Rani, Nanda and Saha, Bikash and Maurya, Vikas and Shukla, Sandeep Kumar}, ``{TTPHunter: Automated Extraction of Actionable Intelligence as TTPs from Narrative Threat Reports},'' \url{https://github.com/nanda-rani/TTPHunter-Automated-Extraction-of-Actionable-Intelligence-as-TTPs-from-Narrative-Threat-Reports}, 2025.

\bibitem{TTPXHunter}
N.~Rani, B.~Saha, V.~Maurya, and S.~K. Shukla, ``Ttpxhunter: Actionable threat intelligence extraction as ttps from finished cyber threat reports,'' \emph{Digital Threats: Research and Practice}, vol.~5, no.~4, pp. 1--19, 2024.

\bibitem{orbinato2022automatic}
V.~Orbinato, M.~Barbaraci, R.~Natella, and D.~Cotroneo, ``Automatic mapping of unstructured cyber threat intelligence: an experimental study:(practical experience report),'' in \emph{2022 IEEE 33rd International Symposium on Software Reliability Engineering (ISSRE)}.\hskip 1em plus 0.5em minus 0.4em\relax IEEE, 2022, pp. 181--192.

\bibitem{dessertlab_cti_to_mitre}
{DessertLab}, ``{cti-to-mitre-with-nlp},'' \url {https://github.com/dessertlab/cti-to-mitre-with-nlp/tree/main}, 2023.

\bibitem{legoy2020rcatt_th}
V.~Legoy, A.~Peter, C.~Seifert, and M.~Caselli, ``{MSc Thesis: rcATT},'' \url{https://github.com/vlegoy/rcATT/blob/master/MScThesis_rcATT_VLegoy.pdf}, 2020.

\bibitem{legoy2020rcatt}
V.~Legoy, ``{rcATT},'' \url {https://github.com/vlegoy/rcATT/tree/master}, 2020.

\bibitem{MITREtrieval}
Y.-T. Huang, R.~Vaitheeshwari, M.-C. Chen, Y.-D. Lin, R.-H. Hwang, P.-C. Lin, Y.-C. Lai, E.~H.-K. Wu, C.-H. Chen, Z.-J. Liao \emph{et~al.}, ``Mitretrieval: Retrieving mitre techniques from unstructured threat reports by fusion of deep learning and ontology,'' \emph{IEEE Transactions on Network and Service Management}, 2024.

\bibitem{wmlab_MITREtrieval}
{WMLab}, ``{MITREtrieval},'' \url{https://github.com/wmlab-MITREtreival/MITREtrieval}, 2023.

\bibitem{mitre2023tram}
{MITRE Engenuity}, ``{Our TRAM: Large Language Model Automates TTP Identification in CTI Reports},'' \url {https://medium.com/mitre-engenuity/our-tram-large-language-model-automates-ttp-identification-in-cti-reports-5bc0a30d4567}, 2023.

\bibitem{mitre_tram_github}
{Center for Threat-Informed Defense}, ``{TRAM},'' \url {https://github.com/center-for-threat-informed-defense/tram}, 2023.

\bibitem{xu2023intelEX}
M.~Xu, H.~Wang, J.~Liu, Y.~Lin, C.~X.~Y. Liu, H.~W. Lim, and J.~S. Dong, ``Intelex: A llm-driven attack-level threat intelligence extraction framework,'' \emph{arXiv preprint arXiv:2412.10872}, 2024.

\bibitem{intelex11dataset}
{Center for Threat-Informed Defense}, ``{IntelEX Dataset},'' \url {https://sites.google.com/view/intelex11/datasets}, 2024.

\bibitem{alam2023looking}
M.~T. Alam, D.~Bhusal, Y.~Park, and N.~Rastogi, ``Looking beyond iocs: Automatically extracting attack patterns from external cti,'' in \emph{Proceedings of the 26th International Symposium on Research in Attacks, Intrusions and Defenses}, 2023, pp. 92--108.

\bibitem{ladder2023}
{Alam, Md Tanvirul and Bhusal, Dipkamal and Park, Youngja and Rastogi, Nidhi}, ``{LADDER: Looking Beyond IoCs Dataset},'' \url{https://github.com/aiforsec/LADDER}, 2023.

\bibitem{schwartz2024llmcloudhunter}
Y.~Schwartz, L.~Benshimol, D.~Mimran, Y.~Elovici, and A.~Shabtai, ``Llmcloudhunter: Harnessing llms for automated extraction of detection rules from cloud-based cti,'' \emph{arXiv preprint arXiv:2407.05194}, 2024.

\bibitem{siracusano2023time}
G.~Siracusano, D.~Sanvito, R.~Gonzalez, M.~Srinivasan, S.~Kamatchi, W.~Takahashi, M.~Kawakita, T.~Kakumaru, and R.~Bifulco, ``Time for action: Automated analysis of cyber threat intelligence in the wild,'' \emph{arXiv preprint arXiv:2307.10214}, 2023.

\bibitem{zhang2024attackgk}
Y.~Zhang, T.~Du, Y.~Ma, X.~Wang, Y.~Xie, G.~Yang, Y.~Lu, and E.-C. Chang, ``Attackg+: Boosting attack knowledge graph construction with large language models,'' \emph{arXiv preprint arXiv:2405.04753}, 2024.

\bibitem{alam2024ctibench}
M.~T. Alam, D.~Bhusal, L.~Nguyen, and N.~Rastogi, ``Ctibench: A benchmark for evaluating llms in cyber threat intelligence,'' \emph{arXiv preprint arXiv:2406.07599}, 2024.

\bibitem{ctibench2024}
{Alam, Md Tanvirul and Bhusal, Dipkamal and Nguyen, Le and Rastogi, Nidhi}, ``{CTIBench: A Benchmark for Evaluating LLMs in Cyber Threat Intelligence},'' \url{https://github.com/xashru/cti-bench/tree/main}, 2024.

\bibitem{ibm_threat_intelligence}
{IBM}, ``{What is threat intelligence?}'' \url{https://www.ibm.com/think/topics/threat-intelligence}.

\bibitem{mitre_attack}
{MITRE Corporation}, ``{MITRE ATT\&CK Framework},'' \url{https://attack.mitre.org/}.

\bibitem{arazzi2023nlp}
M.~Arazzi, D.~R. Arikkat, S.~Nicolazzo, A.~Nocera, M.~Conti \emph{et~al.}, ``Nlp-based techniques for cyber threat intelligence,'' \emph{arXiv preprint arXiv:2311.08807}, 2023.

\bibitem{yao2024llmlieshallucinationsbugs}
J.-Y. Yao, K.-P. Ning, Z.-H. Liu, M.-N. Ning, Y.-Y. Liu, and L.~Yuan, ``Llm lies: Hallucinations are not bugs, but features as adversarial examples,'' \emph{arXiv preprint arXiv:2310.01469}, 2023.

\bibitem{chatgpt_label}
A.~H. Nasution and A.~Onan, ``Chatgpt label: Comparing the quality of human-generated and llm-generated annotations in low-resource language nlp tasks,'' \emph{IEEE Access}, 2024.

\bibitem{chatgpt_vs_human}
M.~Aldeen, J.~Luo, A.~Lian, V.~Zheng, A.~Hong, P.~Yetukuri, and L.~Cheng, ``Chatgpt vs. human annotators: A comprehensive analysis of chatgpt for text annotation,'' in \emph{2023 International Conference on Machine Learning and Applications (ICMLA)}.\hskip 1em plus 0.5em minus 0.4em\relax IEEE, 2023, pp. 602--609.

\bibitem{ctid_adversary_emulation_library}
{Center for Threat-Informed Defense}, ``{Adversary Emulation Library},'' \url{https://github.com/center-for-threat-informed-defense/adversary_emulation_library}.

\bibitem{apt29_emulation_library}
{{Center for Threat-Informed Defense}}, ``{{APT29 Adversary Emulation Library}},'' \url{https://github.com/center-for-threat-informed-defense/adversary_emulation_library/tree/master/apt29}.

\bibitem{mitre_apt29}
{{MITRE ATT{\&}CK}}, ``{{APT29}},'' \url{https://attack.mitre.org/groups/G0016/}.

\bibitem{carbanak_emulation_library}
{{Center for Threat-Informed Defense}}, ``{{Carbanak Adversary Emulation Library}},'' \url{https://github.com/center-for-threat-informed-defense/adversary_emulation_library/tree/master/carbanak}.

\bibitem{mitre_carbanak}
{MITRE}, ``{Carbanak (G0008)},'' \url{https://attack.mitre.org/groups/G0008/}.

\bibitem{fin6_emulation_library}
{{Center for Threat-Informed Defense}}, ``{{FIN6 Adversary Emulation Library}},'' \url{https://github.com/center-for-threat-informed-defense/adversary_emulation_library/tree/master/fin6}.

\bibitem{mitre_fin6}
{MITRE}, ``{FIN6 (G0037)},'' \url{https://attack.mitre.org/groups/G0037/}.

\bibitem{fin7_emulation_library}
{{Center for Threat-Informed Defense}}, ``{{FIN7 Adversary Emulation Library}},'' \url{https://github.com/center-for-threat-informed-defense/adversary_emulation_library/tree/master/fin7}.

\bibitem{mitre_fin7}
{MITRE}, ``{FIN7 (G0046)},'' \url{https://attack.mitre.org/groups/G0046/}.

\bibitem{oilrig_emulation_library}
{{Center for Threat-Informed Defense}}, ``{{OilRig Adversary Emulation Library}},'' \url{https://github.com/center-for-threat-informed-defense/adversary_emulation_library/tree/master/oilrig}.

\bibitem{mitre_oilrig}
{MITRE}, ``{OilRig (G0049)},'' \url{https://attack.mitre.org/groups/G0049/}.

\bibitem{sandworm_emulation_library}
{{Center for Threat-Informed Defense}}, ``{{Sandworm Adversary Emulation Library}},'' \url{https://github.com/center-for-threat-informed-defense/adversary_emulation_library/tree/master/sandworm}.

\bibitem{mitre_sandworm}
{MITRE}, ``{Sandworm (G0034)},'' \url{https://attack.mitre.org/groups/G0034/}.

\bibitem{wizardspider_emulation_library}
{{Center for Threat-Informed Defense}}, ``{{WizardSpider Adversary Emulation Library}},'' \url{https://github.com/center-for-threat-informed-defense/adversary_emulation_library/tree/master/wizardspider}.

\bibitem{mitre_wizardspider}
{MITRE}, ``{Wizard Spider (G0102)},'' \url{https://attack.mitre.org/groups/G0102/}.

\bibitem{crowdstrike_ryuk}
{CrowdStrike}, ``{Big Game Hunting with Ryuk: Another Lucrative Targeted Ransomware},'' \url{https://www.crowdstrike.com/en-us/blog/big-game-hunting-with-ryuk-another-lucrative-targeted-ransomware/}.

\bibitem{mitre_t1059_001}
{MITRE}, ``{Command and Scripting Interpreter - PowerShell (T1059.001)},'' \url{https://attack.mitre.org/techniques/T1059/001/}.

\bibitem{pdfannots_github}
{Abu}, ``{PDFannots: Extracts Annotations from PDF Files},'' \url{https://github.com/0xabu/pdfannots}.

\bibitem{mitre_t1059}
{MITRE}, ``{Command and Scripting Interpreter (T1059)},'' \url{https://attack.mitre.org/techniques/T1059/}.

\bibitem{mitre_t1566}
MITRE, ``{Phishing (T1566)},'' \url{https://attack.mitre.org/techniques/T1566/}.

\bibitem{mitre_t1027}
{MITRE}, ``{Obfuscated Files or Information (T1027)},'' \url{https://attack.mitre.org/techniques/T1027/}.

\bibitem{mitre_t1105}
MITRE, ``{Ingress Tool Transfer (T1105)},'' \url{https://attack.mitre.org/techniques/T1105/}.

\bibitem{mitre_t1071}
{MITRE}, ``{Application Layer Protocol (T1071)},'' \url{https://attack.mitre.org/techniques/T1071/}.

\bibitem{mitre_s0030}
MITRE, ``{Carbanak (S0030)},'' \url{https://attack.mitre.org/software/S0030/}.

\bibitem{mitre_s0046}
{MITRE}, ``{CozyDuke (S0046)},'' \url{https://attack.mitre.org/software/S0046/}.

\bibitem{mitre_s0050}
MITRE, ``{CosmicDuke (S0050)},'' \url{https://attack.mitre.org/software/S0050/}.

\bibitem{mitre_s0154}
{MITRE}, ``{Cobalt Strike (S0154)},'' \url{https://attack.mitre.org/software/S0154/}.

\bibitem{mitre_s0053}
MITRE, ``{SeaDuke (S0053)},'' \url{https://attack.mitre.org/software/S0053/}.

\bibitem{krippendorff2004content}
K.~Krippendorff, \emph{{Content Analysis: An Introduction to Its Methodology}}.\hskip 1em plus 0.5em minus 0.4em\relax Thousand Oaks, CA: Sage Publications, 2004.

\bibitem{krippendorff2013reliability}
{Klaus Krippendorff}, ``{Reliability in Content Analysis},'' \emph{Human Communication Research}, 2013.

\bibitem{python_difflib_sequenceMatcher}
{Python Software Foundation}, ``{difflib -- SequenceMatcher.ratio},'' \url{https://docs.python.org/3/library/difflib.html#difflib.SequenceMatcher.ratio}.

\bibitem{acl_p02_1040}
{K. S. Jones and K. E. Kummerfeld}, ``{The State of the Art in Information Retrieval Evaluation},'' in \emph{{Proceedings of the 2002 Conference on Empirical Methods in Natural Language Processing}}, 2002.

\bibitem{spacy}
{Explosion AI}, ``{spaCy: Industrial-Strength Natural Language Processing in Python},'' \url{https://spacy.io/}.

\bibitem{spacy_model_en}
{{SpaCy Models}}, ``{{SpaCy Models: English Language}},'' \url{https://spacy.io/models/en}.

\bibitem{google2025_apt29_phishing}
{Cloud Security Team}, ``{Not So Cozy: An Uncomfortable Examination of a Suspected APT29 Phishing Campaign},'' \url{https://cloud.google.com/blog/topics/threat-intelligence/not-so-cozy-an-uncomfortable-examination-of-a-suspected-apt29-phishing-campaign/}, 2025.

\bibitem{anthropic2024claude3}
{Anthropic}, ``{Introducing the next generation of Claude},'' \url{https://www.anthropic.com/news/claude-3-family}, 2024.

\bibitem{deepset2024promptengineering}
{Deepset}, ``{Prompt Engineering Guidelines},'' \url{https://docs.cloud.deepset.ai/docs/prompt-engineering-guidelines}, 2024.

\end{thebibliography}

%\input{sections/appendix}

%\theendnotes

\end{document}